\documentclass[sigplan, screen, nonacm]{acmart}

\setcopyright{none}

\usepackage{booktabs}   
\usepackage{subcaption} 

\usepackage{listings}

\usepackage{hoareTriples}
\usepackage{separationLogic}
\usepackage{variables}

\begin{document}

\title
	{%
		Completeness Thresholds for Memory Safety:
		Unbounded Guarantees via Bounded Proofs
	}
\subtitle{(Extended Abstract)}                     


\author{Tobias Reinhard}
	\email{tobias.reinhard@kuleuven.be}
	\orcid{0000-0003-1048-8735}             
	
\affiliation{
		\institution{KU Leuven}            
		\country{Belgium}                    
	}


\author{Justus Fasse}
	\email{justus.fasse@kuleuven.be}
	\orcid{0009-0006-7383-7866}

\affiliation{
		\institution{KU Leuven}            
		\country{Belgium}                    
	}
	
	
\author{Bart Jacobs}
	\email{bart.jacobs@kuleuven.be}
	\orcid{0000-0002-3605-249X}

\affiliation{
  \institution{KU Leuven}            
  \country{Belgium}                    
}



\maketitle


\definecolor{codegreen}{rgb}{0,0.6,0}
\definecolor{codegray}{rgb}{0.5,0.5,0.5}
\definecolor{codepurple}{rgb}{0.58,0,0.82}
\definecolor{backcolour}{rgb}{0.95,0.95,0.92}

\lstdefinestyle{WHILEstyle}{
	backgroundcolor=\color{backcolour},   
	commentstyle=\color{codegreen},
	keywordstyle=\color{magenta},
	numberstyle=\tiny\color{codegray},
	stringstyle=\color{codepurple},
	basicstyle=\ttfamily\footnotesize,
	breakatwhitespace=false,         
	breaklines=true,                 
	captionpos=b,                    
	keepspaces=true,                 
	numbers=left,                    
	numbersep=5pt,                  
	showspaces=false,                
	showstringspaces=false,
	showtabs=false,                  
	tabsize=2
}
\lstset{style=WHILEstyle}

\section{Running Example}

\begin{figure}
\begin{lstlisting}
{ array(a, s) * F }			  // Precondition
  f;          // Complex code not mentioning a, s
  if s > B then
    not_sorted := true;   // bubble_sort(a)
    while not_sorted do					 
      not_sorted := false;
      for i in [L : s-R] do				
        if a[i+1] < a[i] then
          not_sorted := true;
          tmp := a[i];
          a[i] := a[i+1];
          a[i+1] := tmp;  // End bubble_sort(a)
    r := a[Y];              
  g           // Complex code not mentioning a, s
{ array(a, s) * F }       // Postcondition
\end{lstlisting}
	\caption{
		Program \exProgVar sorting array \arrayVar of size \sizeVar.
	}
	\label{fig:code:whole_program}
\end{figure}

Suppose we want to verify that the program \exProgVar
presented in Fig.~\ref{fig:code:whole_program} is memory safe.
The separation logic~\cite{DBLP:journals/cacm/OHearn19,OHearn2001LocalRA,Reynolds2002SeparationLA} precondition $\arrayPred \slStar F$ describes the memory layout we expect at the start: 
An allocated array \arrayVar of size \sizeVar and 
%
%
some disjoint memory $F$ needed for~$f, g$.
%
In line 2, we start with some complex computation~$f$ that does not concern the array \arrayVar nor its size \sizeVar.
In lines 3-12 we bubble sort \arrayVar.
Afterwards, we select the element at index~$Y$ in line 13.
We conclude with some complex code~$g$ in line 14 
that concerns neither \arrayVar nor \sizeVar.
The postcondition in line 15 states that the memory layout remains unchanged.
$B, L, R, Y \geq 0$ are positive constants.

\section{Unbounded vs Bounded Proofs}
We can choose between two different approaches to verify \exProgVar:
Unbounded and bounded proofs.
An unbounded proof ensures that the program is safe for all possible inputs, in particular for all array sizes \sizeVar, i.e.,
$\models 
	\forall \sizeVar.\ 
	\hoareTriple
		{\arrayPred \slStar F}
		{\exProgVar}
		{\arrayPred \slStar F}
$.
Such strong guarantees come, however, typically at the cost of writing tedious inductive proofs.

A bounded proof on the other hand only considers inputs up to a typically small size bound, e.g., $\sizeVar < 10$.
Hence, it only yields bounded guarantees, i.e.,
$\models 
	\forall \sizeVar < 10.\ 
	\hoareTriple
		{\arrayPred \slStar F}
		{\exProgVar}
		{\arrayPred \slStar F}
$.
The big advantage is that bounded proofs are easy to automate.
We often justify them with the intuition that bugs tend to show up for small inputs.
Since this intuition is not based on any formal guarantees, bounded proofs can easily convey a false sense of correctness.

\section{Completeness Thresholds}
The connection between bounded and unbounded proofs has been extensively studied for finite state transition systems~\cite{Biere1999SymbolicMC, Clarke2004CompletenessAC, Kroening2003EfficientCO, Bundala2012OnTM, Abdulaziz2018FormallyVA, Heljanko2005IncrementalAC, Awedh2004ProvingMP, McMillan2003InterpolationAS}.
For a system~$T$ and target property $\phi$, a \emph{completeness threshold} (CT) is any number $k$ such that examining path prefixes of length $k$ is sufficient to prove $\phi$, i.e.,
$T \models_k \phi \Rightarrow T \models \phi$~\cite{Clarke2004CompletenessAC}.
The CTs described in the literature are characterised in terms of finite key properties of the transition system $T$, such as the recurrence diameter~\cite{Kroening2003EfficientCO}.

Programs like \exProgVar that process arbitrary big data correspond to infinite trasition systems, for which these key attributes are in general infinite.
Hence, we cannot reuse existing definitions and results from the model checking literature.
In~\cite{Reinhard2023SOAP-ct4ms} we propose a new definition of CTs for program specifications and show that we can extract them by analysing a program's verification condition~\cite{Flanagan2001AvoidingExpExplosionVC, Parthasarathy2021VCGenerator}.

\begin{definition}[Completeness Thresholds for Programs]
	Let \hoareTriple{\preVar}{\cmdVar}{\postVar} be a program specification containing a free variable $x$ with domain \domVar.
	We call a subdomain $\ctVar \subseteq \domVar$ a \emph{completeness threshold} for $x$ in \hoareTriple{\preVar}{\cmdVar}{\postVar} if
	$$
		\models 
			\forall x \in \ctVar.\
				\hoareTriple{\preVar}{\cmdVar}{\postVar}
		\quad\Rightarrow\quad
		\models
			\forall x \in \domVar.\
				\hoareTriple{\preVar}{\cmdVar}{\postVar}
	$$
\end{definition}

A subdomain $\ctVar \subseteq \domVar$ is a CT if it is big enough to ensure that a bounded proof only considering \ctVar does not miss bugs.
%
In other words, for every error, \ctVar must contain an input $\varVar \in \domVar$ that allows us to reach it.
%
In the following, we sketch the extraction of a CT for the array size \sizeVar in \exProgVar.



\paragraph{Program Slicing}
We can use program slicing~\cite{Weiser84ProgramSlicing, MastroeniDataDependencyProgramSlicing, Asavoae2018ChiselProgramSlicing, Moser90DataDependencyGraphs} to isolate the parts of \exProgVar whose memory safety is affected by \sizeVar.
This allows us to get rid of the complex code $f$ and $g$ since they mention neither \sizeVar nor \arrayVar.
We can also eliminate the outer bubble sort loop, its condition variable and~$r$.
The remaining program does not mention the memory described by $F$, so we can eliminate it as well.
What remains are the parts of \exProgVar that are potentially affected by the value of \sizeVar.
Hence, program slicing allows us to reduce CTs for \sizeVar in \exProgVar to CTs in the \exProgRedVar presented in Fig.~\ref{fig:code:reduced_program}.
In other words, any CT that we find for \sizeVar in \exProgRedVar is also a CT for \sizeVar in \exProgVar.

\begin{figure}
\begin{lstlisting}
{ array(a, s) }
  if s > B then                // (iv) conditional 
    for i in [L : s-R] do      // (iii) iterator
      if a[i+1] < a[i] then    // (ii) body
        tmp := a[i];           //       |
        a[i] := a[i+1];        //       |
        a[i+1] := tmp;         // ______|
    a[Y];                      // (i)
{ array(a, s) }
\end{lstlisting}
	\caption{
		Reduced program \exProgRedVar.
		$L, R, B$ are constants.
	}
	\label{fig:code:reduced_program}
\end{figure}

\paragraph{Follow the AST}
Memory safety CTs compose well.
Therefore, we follow the structure of the AST when extracting CTs.
We split the slice into four parts and analyse them bottom up:
(i)~The final read in line 8,
(ii)~the loop body in lines 4-7,
(iii)~the iterator in line 3 and
(iv)~the conditional in line 2.

\paragraph{Constraints}
To simplify the notation, we use constraint sets $\{k_1, \dots, k_n\}$ to describe CTs.
Each $k_i$ describes a property that some element in our CT must cover to reach a specific potential error.
If $k_i$ is unsatisfiable, it means that the error is unreachable.
Hence, a set \ctVar models the constraint set if it contains models $\ctElemVar_1, \dots, \ctElemVar_n$ for each satisfiable constraint, i.e.,
$
	\not\models \neg k_i 
	\Rightarrow\,
	\models k_i[\sizeVar \mapsto \ctElemVar_i]
$.

\paragraph{Constant Array Access} 
\textbf{(i)}~Selecting the element at index $Y$ is safe if the array has at least $Y+1$ elements.
In order to catch a potential out-of-bounds error, a bounded proof must consider smaller arrays.
That is, for any $\ctElemVar \leq Y$, the singleton $\{\ctElemVar\}$ is a CT for \sizeVar in $\arrayVar[Y]$.
We get the constraint set 
$\{ \sizeVar \leq Y \}$.

\paragraph{Iterating over Arrays}
\textbf{(ii)}~The loop body contains array accesses $\arrayVar[i], \arrayVar[i+1]$, guarded by the comparison $a[i+1] < a[i]$.
The guard does not depend on \sizeVar, so we can ignore it.
When we study the body's verification condition, we see that it is subsumed by the unguarded array access $\arrayVar[i]; \arrayVar[i+1]$.
In \cite{Reinhard2023SOAP-ct4ms} we showed that in combination with the iterator (iii), the offset does not affect the CT for \sizeVar.
That is, we can reduce the body (ii) to $\arrayVar[i+z]$.
\textbf{(iii)}~In \cite{Reinhard2023SOAP-ct4ms} we showed that any size $\sizeVar \geq L + R$ is a CT for 
\lstinline|for i in [L : s-R] do a[i+z]|.

\paragraph{Conditional and Sequence}
\textbf{(iv)}~A bounded proof for the conditional's body, i.e., lines 3 - 8, reaches all potential errors if it reaches all errors in the loop (iii)+(ii) and in the final read~(i).
We can compute a CT for both by merging their respective constraint sets, i.e., $\{ \sizeVar \geq L + R,\ \, \sizeVar \leq Y \}$.
A bounded proof needs to pass the guard~(iv) in order to check the code on lines 4-8.
We have to refine our constraints accordingly:
$\{ 
	\sizeVar > B \wedge \sizeVar \geq L + R,\ \,
 	\sizeVar > B \wedge \sizeVar \leq Y 
\}$.

Earlier we sliced \exProgVar to identify the parts that are affected by \sizeVar and thereby reduced CTs for \sizeVar in \exProgVar to CTs in \exProgRedVar.
So, we found a CT for \sizeVar in \exProgVar, i.e., any set \ctVar containing $\ctElemVar_1, \ctElemVar_2$ 
with
$\ctElemVar_1 > B \wedge \ctElemVar_1 \geq L + R$
and
$\ctElemVar_2 > B \wedge \ctElemVar_2 \leq Y$.
Though, the concrete models depend on the constants $B,L,R,Y$.

\paragraph{Increase Trust in Bounded Model Checking}
Suppose the complex code $f, g$ in \exProgVar involves a list~$l$ of size~$n$.
Any bounded proof for \exProgVar has to bound the sizes \sizeVar and $n$.
The CT for \sizeVar only allows us to derive unbounded guarantees about the code that corncerns \arrayVar and \sizeVar.
The overall guarantees for \exProgVar, however, remain bounded, due to the $n$-bound.

Even if we accept the idea of bounded guarantees, the sizes that we are able to check is a very limiting factor of bounded model checking.
The approach suffers from the state space explosion problem~\cite{Clarke2008BmcStateExplosion, Park2000JavaMC, Clarke2000CounterExGuidedRefinement}.
Therefore, it is often only practical to check very small bounds, which leads to proofs that convey little trust.

For simplicity, suppose both sizes \sizeVar and $n$ affect the proof complexity to a similar degree.
Suppose we are able to check $\sizeVar, n \leq 10$ within our given resource constraints.
The CT we extracted depends on the constants $B,L,R,Y$.
Let us consider a simple scenario for which our bubble sort implementation is safe:
$B = 1, L = 0, R = 2, Y = 0$.
With these constants, our second constraint becomes unsatisfiable and
$\{2\}$ is a model of our constraint set and hence a CT for \sizeVar.

In other words, our bounded proof is allowed to fix \sizeVar without decreasing the guarantees we get.
A proof for $\sizeVar = 2, n \leq 10$ runs much faster than a proof for $\sizeVar, n \leq 10$.
We can reinvest the freed up time into checking bigger list sizes, e.g., $\sizeVar = 2, n \leq 15$.
Thereby we get a proof with a similar runtime that conveys stronger guarantees.
Hence, it is much more likely that we can trust the result.

\bibliography{bib}

\end{document}